\documentclass{article}

\usepackage{arxiv}

\usepackage[utf8]{inputenc} 
\usepackage[T1]{fontenc}    
\usepackage{hyperref}       
\usepackage{url}            
\usepackage{booktabs}       
\usepackage{amsfonts}       
\usepackage{nicefrac}       
\usepackage{microtype}      
\usepackage{lipsum}
\usepackage{bm}
\usepackage{graphicx}
\usepackage{enumerate}
\usepackage{siunitx}
\usepackage[]{algorithm2e}
\usepackage{amsmath}

\title{Interface Reconstruction and Advection schemes for Volume of fluid method in Axisymmetric coordinates}

\author{
  Ananthan M\\
  Department of Mechanical Engineering\\
  Indian Institute of Science\\
  Bangalore, India \\
  \texttt{ananthanm@iisc.ac.in} \\
   \And
  Gaurav Tomar\thanks{Corresponding author.} \\
  Department of Mechanical Engineering\\
  Indian Institute of Science\\
  Bangalore, India \\
  \texttt{gtom@iisc.ac.in} \\
}

\begin{document}
\maketitle

\begin{abstract}

Volume of fluid(VOF) method is a sharp interface method employed for simulations of two phase flows. Interface in VOF is usually represented
 using piecewise linear line segments in each computational grid based on the volume fraction field. While VOF for cartesian
 coordinates conserve mass exactly, existing algorithms do not show machine-precision mass conservation for axisymmetric
 coordinate systems. In this work, we propose analytic formulae for interface reconstruction in axisymmetric coordinates,
 similar to those proposed by Scardovelli and Zaleski (J. Comput. Phys. 2000) for cartesian coordinates. We also propose modifications to the
 existing advection schemes in VOF for axisymmetric coordinates to obtain higher accuracy in mass conservation. 

\end{abstract}


\section{Introduction}

Multiphase flows are ubiquitious in several industrial applications. In the last three decades, there has been a surge
 in the numerical methods and algorithms for simulations of complex multiphase flows. There have been several different types of
 interface capturing strategies that have been proposed for two-phase flows. The most popular of these are the Level set method,
  Volume of Fluid (VOF) method, and Front tracking scheme \cite{ProsperettiTryggvasonBook,TryggvasonZeleskiBook}. VOF methods
 with geometric advection strictly conserve the volume of the two phases.

Several improvements have been made since the inception of the method (see Hirt and Nichols\cite{Hirt1981,Renardy2002,Youngs1982,
Pilliod1992,Pilliod2004,Scardovelli2003}). The most simplest and earliest representation for interface reconstruction is simple
 line interface calculation(SLIC) in which the interface is approximated by
 horizontal or vertical lines. Subsequently, piecewise line interface construction (PLIC) was introduced where the interface
 is approximated as a linear line at an angle in the cell \cite{Youngs1984}. Higher order interface construction have been
 proposed (such as Parabolic reconstruction by \cite{Renardy2002}), but considering the associated computational cost and
 complexity for geometric advection,  PLIC is usually preferred.

Scradovilli and Zaleski\cite{Scardovelli2000} proposed analytical formulae for the piecewise linear reconstruction of the interface
 in cartesian coordinates that led to a significant speedup over the earlier iterative schemes. However, for curvilinear
 coordinates (such as axisymmetric coordinate system), the proposed analytical formulae cannot be employed directly. In the
 present work, we derive similar analytic formulae for axisymmetric coordinates, which result in a speedup of $\sim 28$ over the iterative counterparts
 (Brent's root finding method). Further, we demonstrate that the existing interface advection schemes in VOF for axisymmetric
 coordinates are not strictly mass conserving. In this study, we propose modifications to the current operator split
 algorithms that result in machine-precision mass conservation in axisymmetric coordinates. We show the efficacy of the proposed algorithms using several test cases.

The paper is organized as follows. We first present analytical formulae for the interface reconstruction schemes in axisymmetric
 coordinates in section 2. In section 3, we propose modifications in the existing interface advection algorithm for axisymmetric
 VOF and present test cases to show the efficacy of the scheme. Finally, in section 4, we discuss the important conclusions.

\section{Interface Reconstruction Scheme}

\begin{figure}[htbp]
    \centerline{\includegraphics[width=2.5in]{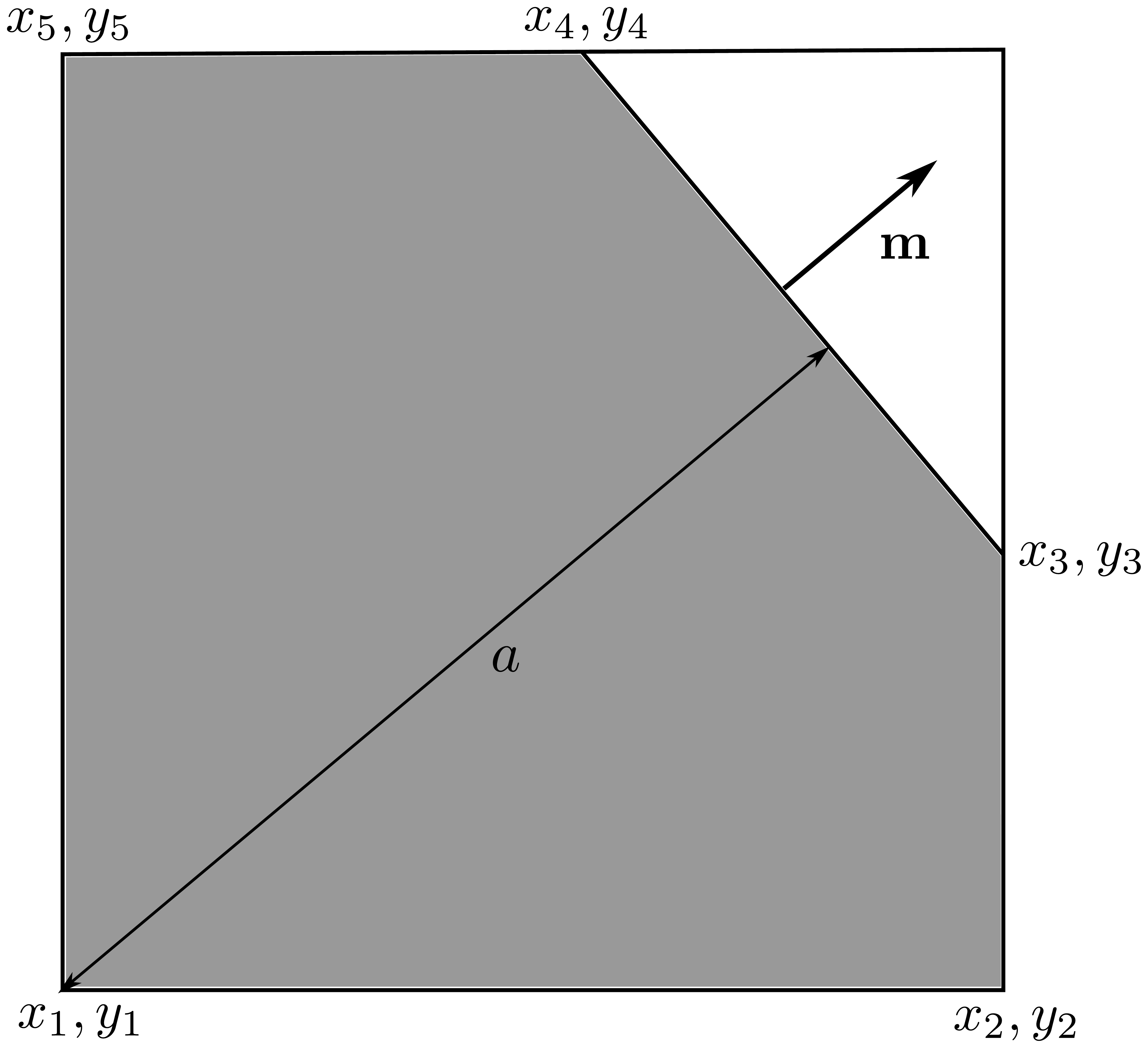}}
\caption{Coordinates of the vertices of a simple polygon cut by the interface ordered counter clockwise. Here $x_3,y_3$ and
    $x_4,y_4$ represents the end points of the PLIC line segment with interface normal $\bold{m}$ and line constant $a$. }
\label{fig:shoelace}
\end{figure}

Interface reconstruction in the volume of fluid(VOF) method requires the volume fraction field. Using the volume fraction
field, a piece-wise linear or a higher order interface is constructed in a given grid-cell.
 Interface reconstruction is an integral part of the geometric advection schemes to ensure mass conservation property of the VOF
 method \cite{TryggvasonZeleskiBook}. Initial condition for a multiphase flow simulation requires the initial distribution of the
 volume fraction field, usually provided as an implicit function of the spatial coordinates.  VOFI library\cite{Bna2016} is an
open source library to initialise the liquid volume fraction field in cartesian coordinate systems accurately. In VOFI, for cells
 cut by the interface (see figure \ref{fig:shoelace}), PLIC reconstruction method \cite{Youngs1984,Pilliod1992} is employed to approximate
 the interface as a linear line segment,
\begin{equation}
  \boldsymbol{m}.\boldsymbol{x}  = a,
  \label{eq:planeeqn}
\end{equation}
where $\boldsymbol{m}$ is the local normal at the interface, $\boldsymbol{x}$ is a point on the plane, and $a$ is the normal
 distance of the origin from the plane. Analytical relation, given by Scardovelli and Zaleski\cite{Scardovelli2000}, between
 the volume fraction and the line constant is employed to determine the line constant $a$. Thus, for two-dimensional and
 three-dimensional Cartesian coordinate systems, VOFI library can be directly employed for accurate assignment of the initial
 volume fraction field on a given discretized domain using an implicit equation of the interface. However, in curvilinear coordinate
 systems, for a given implicit function, the piece-wise linear interface constructed from VOFI would require computation of the volume
 fraction field using a formula specific to the curvilinear coordinates. For instance, for axisymmetric coordinate system, the
 modified Gauss area (shoelace) formula  for computation of the area of a convex polygon is given by,
\begin{equation}
V = \frac{\pi}{3}\left|\sum_{i=1}^{n}\left( x_{i} + x_{i+1} \right)\left(x_{i} y_{i+1}-x_{i+1} y_{i}\right)\right|
\label{formula_shoelace}
\end{equation}
where $(x_i,y_i)$ for $i = 1,...,n$ (with $x_{n+1}=x_{1}$ and $y_{n+1}=y_{1}$) are the coordinates of the vertices of a
 convex polygon ordered counter clockwise as shown in the figure \ref{fig:shoelace}.

Thus, for initialization of the volume fraction field, $C$, once we obtain the linear interface in each grid cell using the VOFI
 library, we use the above formula to compute the volume, $V$, and assign volume fraction in each grid cell as,
\begin{equation}
  C = \frac{V}{2 \pi r_c \Delta x \Delta y}.
  \label{eq:vol-fraction}
\end{equation}
Here $r_c$ is the distance of the center of the cell from the axis of symmetry, and $\Delta x$ and $\Delta y$ are the grid-cell
sizes in the radial ($r$) and axial ($y$) directions, respectively. We note that the above procedure is followed essentially to minimize the error in
the  volume-fraction during initialization. To illustrate this, we initialise a torus of minor radius $r_t=0.25$ and major radius $r = 0.50$
 in the center of a computational domain of size $1\times1$ as shown in the figure \ref{Test1}. The volume of the torus can be
 analytically computed as $V_t=2 \pi^2 r r_t^2$, where the major radius, $r$, is the distance to the center of the torus from
 the axis of symmetry. We compare the results for various grid sizes with the results obtained using the popular VOF based open source flow solver,
  Gerris\cite{popinet2003}, given in table \ref{table1}.

\begin{table}
    \caption{Results for relative error in volume during initialization of a torus of radius, $r_t = 0.25$, for different grid sizes.}
  \centering
  \begin{tabular}{lll}
    \toprule
    \multicolumn{3}{c}{Relative error in volume: $E = \frac{\left|V - V_t \right|}{V_t}$}                   \\
    \cmidrule(r){1-3}
 Grid& Current Solver&Gerris Solver\\
    \midrule
    $16 \times 16$   & $ \num{5.4e-16}$    & $\num{9.4e-3}$\\
    $32 \times 32$   & $ \num{7.1e-16}$    & $\num{2.6e-3}$\\
    $64 \times 64$   & $ \num{1.8e-16}$    & $\num{5.5e-4}$\\
    $128 \times 128$   & $\num{5.4e-16}$    & $\num{1.6e-4}$\\
    \bottomrule
  \end{tabular}
    \label{table1}
\end{table}


\begin{figure}[!htbp]
    \centerline{\includegraphics[width=2.5in]{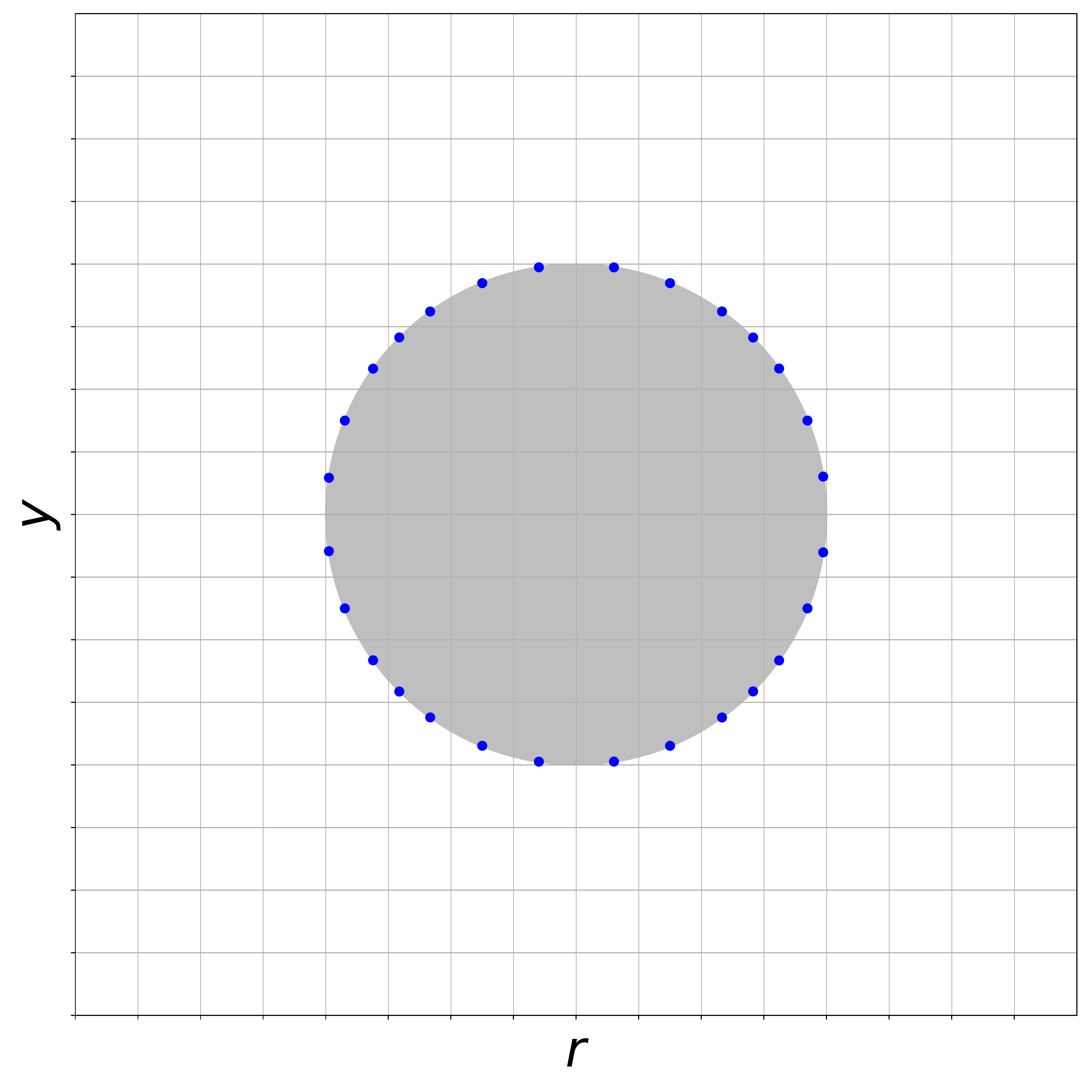}}
    \caption{Reconstructed interface for 16 mesh points of a torus of radius $r_t = 0.25$ and major axis $r = 0.50$ initialised at the center of $1\times1$ domain 
    where the dots represent the mid–points of the reconstructed PLIC line segments.}
    \label{Test1}
\end{figure}

Thus, we have shown that for curvilinear coordinates volume fraction field can be initialized up to machine accuracy. Now, we 
derive an analytic relation for PLIC reconstruction for axisymmetric coordinates on the lines of Scardovelli and Zaleski
\cite{Scardovelli2000}. We use Youngs method\cite{Youngs1984} to get the interface normal ($\boldsymbol{m}$ in 
equation.\ref{eq:planeeqn}) from fluid-1 ($C = 1$) to fluid-2 $C = 0$ and is given by $\boldsymbol{m}= -\nabla C /|\nabla C| $. To
complete the PLIC interface reconstruction for a given $C$, in addition to the normal $\boldsymbol{m}$, we also need to obtain
the line constant, $a$, which is the normal distance of the interface from one of the vertices of the computational cell. In 
what follows, we present a methodology to get the line constant($a$) analytically for a given interface normal vector and the
volume  fraction of a mixed cell.

\begin{figure}[!htbp]
    \centerline{\includegraphics[width=16.45cm]{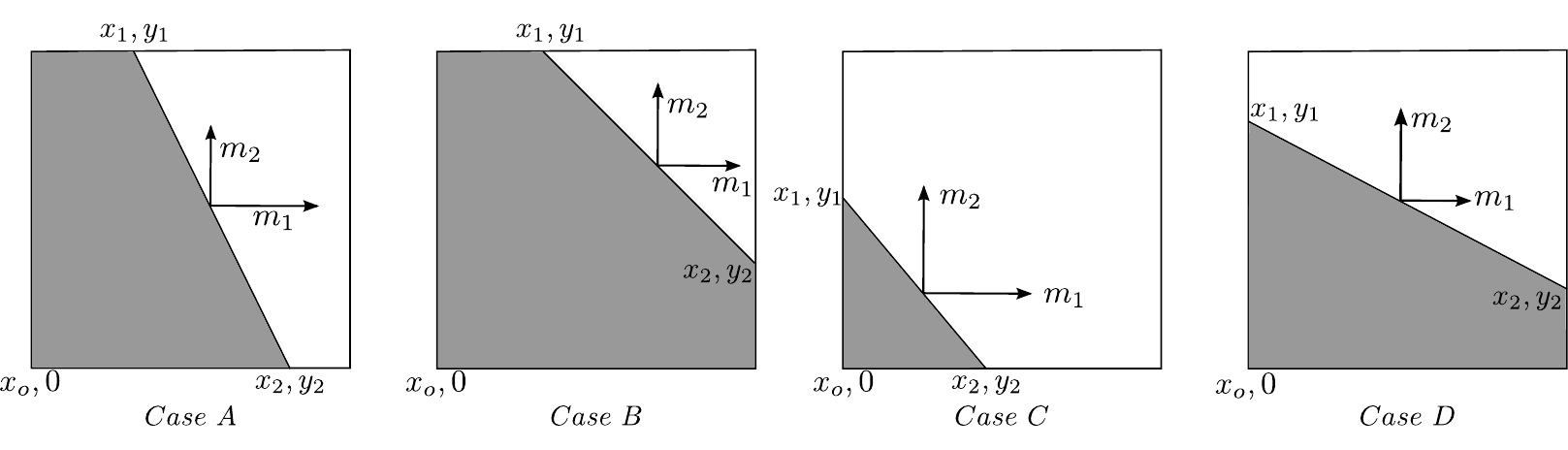}}
    \caption{Various cases which can arise in the standard configuration of the interface with $m_1,m_2 \geq 0$ and fluid $1$ is occupying
     the bottom left corner of the cell.}
    \label{fig:interface-configs}
\end{figure}

As discussed in \cite{Scardovelli2000}, using an analytical relation between the volume fraction($C$), interface normal
 ($\boldsymbol{m}$) and the line constant ($a$), we can implement an $if-else-if-end-if$ construct to determine the line constant, $a$.
 This approach is computationally much more efficient compared to the alternative iterative approach to get the line constant.
Given the equation of the interface, $m_1 x + m_2 y = a$, all combinations of $m_1, m_2$ (such that $m_1^2 + m_2^2 = 1$) can be
 reduced to one of the cases shown in figure \ref{fig:interface-configs}, either by changing the origin or by changing the
 reference fluid from fluid-$1$ to fluid-$2$, such that both $m_1$ and $m_2$ are positive and the left bottom corner of the
 mixed cell under consideration is contained in fluid-$1$. Figure \ref{fig:interface-configs}  shows all the possible
 configurations for interface arrangement with $m_1 \ge 0$ and $m_2 \ge 0$.

We first discuss Case A shown in the figure \ref{fig:interface-configs}  and similar procedure can be followed to obtain relations
 for other cases. For the axisymmetric coordinate system, the shaded area shown in the figure \ref{fig:interface-configs}  for
 Case A is given by,
\begin{equation}
V=\frac{\pi}{3}\left(x_{0}+x_{2}\right)^{3} \frac{m_{1}}{m_{2}}-\frac{\pi}{3}\left(x_{0}+x_{1}\right)^{3} \frac{m_{1}}{m_{2}}-\pi x_{0}^{2} y_{1}.
    \label{VolumeEqn}
\end{equation}

Using the equation of line $ m_1 x + m_2 y = a$ we have $ x_1 = a/m_1 - (m_2 \Delta y)/m_1 $ and $ x_2 = a/m_1 $. In the present
 study, we assume $\Delta x = \Delta y$, but the same analysis can be easily extended for $\Delta x \ne \Delta y$. Substituting
 $x_1$ and $x_2$ in the equation \ref{VolumeEqn} and collecting the terms in powers of $a$, we obtain,
\begin{equation}
\left(\frac{\pi \Delta y}{m_{1}^{2}}\right) a^{2}+\left(-\frac{\pi m_{2} \Delta y^{2}}{m_{1}^{2}}+\frac{2 \pi \Delta y x_{0}}{m_{1}}\right) a+\left(\frac{\pi m_{2}^{2} \Delta y^{3}}{3 m_{1}^{2}}-\frac{\pi \Delta y^{2} m_{2} x_{0}}{m_{1}}\right)=V.
    \label{QuadraticEqn}
\end{equation}
Thus, we have an analytical relation between the volume and the line constant $a$. We note that the above relation holds true only
 when the interface cuts through the top and the bottom edges of the cell shown for Case A in Figure \ref{fig:interface-configs}:
 $(x_{2} - x_{0}) \leq \Delta x$, $y_{1} = \Delta y$ and $y_{2} = 0$. These conditions yield the bounds on the values of
 $a$: $m_{2} \Delta y \le a \leq m_{1} \Delta x$. Substituting the above bounds for $a$ in the equation \ref{QuadraticEqn}, 
yield the bounds on the limiting volumes:
\begin{equation}
V_{1}=\frac{\pi \Delta y\left(3 \Delta x^{2} m_{1}^{2}+3 \Delta x m_{1}\left(2 m_{1} x_{0}-\Delta y m_{2}\right)+\Delta y m_{2}\left(\Delta y m_{2}-3 m_{1} x_{0}\right)\right)}{3 m_{1}^{2}}.
\end{equation}
and
\begin{equation}
    V_{2}=\frac{\pi \Delta y^{2} m_{2}\left(\Delta y m_{2}+3 m_{1} x_{0}\right)}{3 m_{1}^{2}}
\end{equation}
For a given volume fraction $C$ and interface normal ($m_1, m_2$), the volume occupied by fluid-$1$ in the configurations shown
 in the Fig.\ref{fig:interface-configs} is given by: $V = 2\pi r \Delta x \Delta y C$ where $r = x_0 + \Delta x/2$ is the
 distance from the axis of symmetry to the cell center. If $ V_1 \le V \le V_2$ then the analytical relation given by
 equation \ref{QuadraticEqn} can be used to determine the line constant $a$. For the quadratic equation in $a$ given by
 equation. \ref{QuadraticEqn}, we note that only one of the roots will satisfy the required bounds on $a$ for case A. For
 cases B and C, we obtain cubic equations that can be solved for $a$ using Cardano's formula or using Brent's method to find the
 appropriate root with necessary bounds for the line constant. Following the same approach we can get bounds on volume for
 case D as:
\begin{equation}
V_{3}=\frac{\pi \Delta y^{2}\left(-2 \Delta y m_{1}+3 \Delta y m_{2}-3 m_{1} x_{0}+6 m_{2} x_{0}\right)}{3 m_{2}}
\end{equation}

and

\begin{equation}
V_{4}=\frac{\pi \Delta y^{2} m_{1}\left(\Delta y+3 x_{0}\right)}{3 m_{2}}.
\end{equation}

\begin{figure}[!htbp]
    \centerline{\includegraphics[width=5.in]{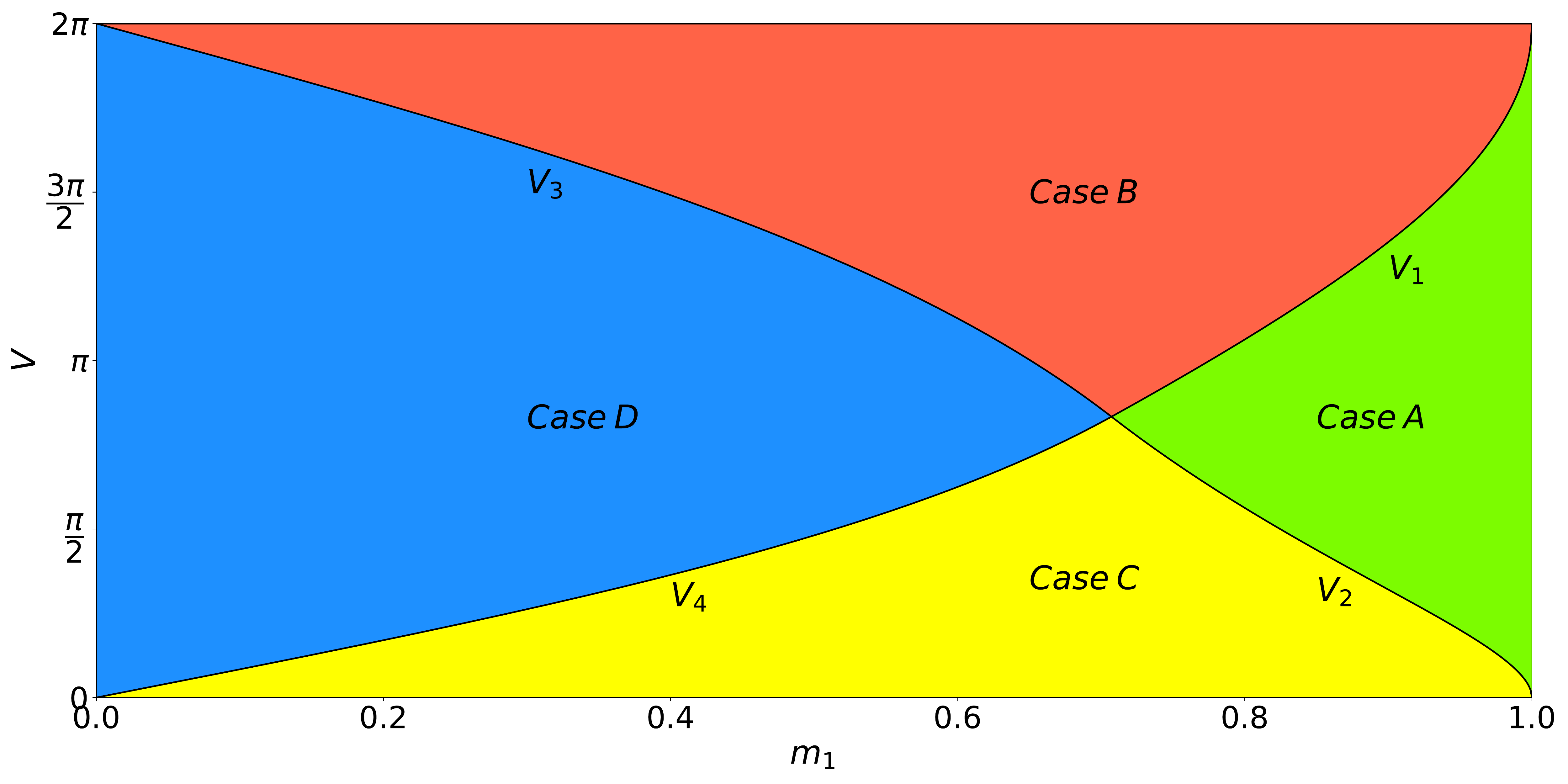}}
    \caption{Bounds for different cases shown in figure \ref{fig:interface-configs} as a function of the x-component of the interface normal, $m_1$. 
    The interfacial cell is placed at a distance of $1$ from the axis of symmetry with $\Delta x = \Delta y =1$. The total volume of the cell is $V_{cell} = 2 \pi$ given by the top boundary in the plot.}
    \label{fig:phasediagram}
\end{figure}
Figure \ref{fig:phasediagram} shows all the possible configurations of the interface and the volume bounds which separate each
case as the interface normal in radial direction varies from minimum to a maximum. Figure \ref{fig:phasediagram} clearly shows that the various bounds for the cases shown in figure \ref{fig:interface-configs} do not overlap and provide a unique criterion for computing the line constant $a$. Thus, we can use the following algorithm
to classify each case.

\begin{algorithm}[H]
\KwData{$V,m_1$}
\KwResult{Identification of the case in which the standard interface belongs to.}
 
\eIf{$\: m_1\geq\frac{1}{\sqrt{2}} \:$}
{
       \uIf{$\:V \geq V_1\:$}
           {
             $Case\:B$
           } 
       \uElseIf{$\:V \leq V_2\:$}
           {
             $Case\:C$
           }
       \Else
           {
             $Case\:A$ \tcp*[f]{$\:V_1 > V > V_2\:$}
           } 
}
{
       \uIf{$\:V \geq V_3\:$}
           {
             $Case\:B$
           } 
       \uElseIf{$\:V \leq V_4\:$}
           {
             $Case\:C$
           }
       \Else
           {
             $Case\:D$ \tcp*[f]{$\:V_3 > V > V_4\:$}
           } 
}
 \caption{Classification of the standard case of the reconstructed interface where $m_1,m_2\geq0$.}
\end{algorithm}

We list below the analytical relation between the line constant($a$) and the volume($V$) for each case:

\subsection*{Case $A$}

\begin{equation}
\left(\frac{\pi \Delta y}{m_{1}^{2}}\right) a^{2}+\left(-\frac{\pi m_{2} \Delta y^{2}}{m_{1}^{2}}+\frac{2 \pi \Delta y x_{0}}{m_{1}}\right) a+\left(\frac{\pi m_{2}^{2} \Delta y^{3}}{3 m_{1}^{2}}-\frac{\pi \Delta y^{2} m_{2} x_{0}}{m_{1}}\right)=V.
\end{equation}

\subsection*{Case $B$}

\begin{equation}
\begin{aligned}
    -\left(\frac{\pi}{3 m_{1}^{2} m_{2}}\right) a^{3}+\left(\frac{\pi \Delta y}{m_{1}^{2}}-\frac{\pi x_{0}}{m_{1} m_{2}}\right) a^{2}+\left(\frac{\pi\left(\Delta y+x_{0}\right)^{2}}{m_{2}}-\frac{\pi m_{2} \Delta y^{2}}{m_{1}^{2}}+\frac{2 \pi \Delta y x_{0}}{m_{1}}-\frac{\pi x_{0}^{2}}{m_{2}}\right) a  \\ 
    +\left(-\frac{\pi \Delta y\left(\Delta y+x_{0}\right)^{2} m_{1}}{m_{2}}+\frac{\pi m_{2}^{2} \Delta y^{3}}{3 m_{1}^{2}}-\frac{\pi \Delta y^{2} m_{2} x_{0}}{m_{1}}-\frac{\pi m_{1} x_{0}^{3}}{3 m_{2}}+\frac{\pi m_{1}\left(\Delta y+x_{0}\right)^{3}}{3 m_{2}}\right)=V
\end{aligned}
\end{equation}

\subsection*{Case $C$}

\begin{equation}
\left(\frac{\pi}{3 m_{1}^{2} m_{2}}\right) a^{3}+\left(\frac{\pi x_{0}}{m_{1} m_{2}}\right) a^{2}=V
\end{equation}

\subsection*{Case $D$}

\begin{equation}
    a=\frac{2 \pi \Delta y^{3} m_{1}+3 \pi \Delta y^{2} m_{1} x_{0}+3 m_{2} V}{3 \pi \Delta y\left(\Delta y+2 x_{0}\right)}
\end{equation}

We note here that the other cases can be readily transformed into one of the cases listed in the figure \ref{fig:interface-configs}
 by either changing the fluid (by using ($1 - C$) instead of $C$ to compute the volume and inverting the interface normal $\boldsymbol{m}$) or by
 changing the origin (keeping the location of the axis-of-symmetry same but inverting its direction).

We now compare the above described analytical method with the iterative method for finding the line constant for the case A
 given in figure \ref{fig:interface-configs}. The relative error in the line constant is given in table \ref{table2} for the
 analytical and iterative methods with different tolerances. We note that the iterative method to reach an error with a
 tolerance of $10^{-8}$ is about $28$ slower compared to the analytical method. 

\begin{table}[h]
    \caption{Relative error in line constant $a$ and the ratio of CPU time required by iterative method(Brent's algorithm) to that required
    by analytical method for $10000$ repetitions for case A shown in figure \ref{fig:interface-configs} for different tolerances used in the iterative method.}
  \centering
  \begin{tabular}{llll}
    \toprule
    \multicolumn{4}{c}{Comparision between the analytical and iterative reconstruction methods}                   \\
    \cmidrule(r){1-4}
    Method& Tolerance&Relative Error&$t_{iterative}/t_{analytical}$\\
    \midrule
    Analytical   &       $ -$&  $ 0 $    & $ 1$\\
    Iterative   &      $ \num{1.0e-4}$&  $ \num{1.8e-4}$    & $\num{14.2}$\\
    Iterative   &      $ \num{1.0e-6}$&  $ \num{1.1e-6}$    & $\num{21.2}$\\
    Iterative   &      $ \num{1.0e-8}$&  $ \num{6.7e-11}$    & $\num{28.3}$\\
    \bottomrule
  \end{tabular}
    \label{table2}
\end{table}

Once the line constant, $a$, is obtained, the position of the endpoints of the linear approximation of the interface can be
 computed thus completing the construction of a planar interface in a given computational cell. As discussed earlier, this more precise description of the interface
  within the grid cell allows geometric advection which gives the VOF method its strict mass conservation property while maintaining a sharp interface.

 In what follows, we discuss an operator split algorithm for the geometric advection of the interface in axisymmetric coordinates.
 We note that the straightforward extension of the 2D cartesian algorithm does not yield accurate results, as also indicated by the
 results obtained from the existing open source codes.

\section{Advection of the Interface}

 We present here a scheme for accurate geometric advection of the volume fraction in the axisymmetric coordinates. We have used
 a uniform grid to describe the variables with volume fraction being stored at the cell centers ($C_{i,j}$). The incompressible
 fluid flow is determined by the velocity field which is defined at the cell faces ($u_{i+1/2,j},v_{i,j+1/2}$). Here, $u$ denotes
 the radial direction velocity and $v$ is the axial velocity. The velocity field satisfies the discrete divergence free condition
 given by:
\begin{equation}
\frac{(r u)_{i+\frac{1}{2}, j}-(r u)_{i-\frac{1}{2}, j}}{r_{i} \Delta x}+\frac{v_{i, j+\frac{1}{2}}-v_{i, j-\frac{1}{2}}}{\Delta y} = 0.
\end{equation}

Motion of the interface is governed by the advection equation for the volume fraction field,
\begin{equation}
\frac{\partial C}{\partial t}+ \mathbf{u} \cdot \nabla C = 0
    \label{adveqn1}
\end{equation}

For incompressible fluids, conservation of the individual volumes of the two fluids results in the conservation of mass. Thus,
 in the volume of fluid method, geometric advection of the volume fraction field is expected to yield machine-precision mass
 conservation. Given a volume fraction field, reconstructed interface and solenoidal velocity field, we can solve the equation
 \ref{adveqn1} using an operator splitting algorithm consisting of an $x-$sweep and a $y-$sweep following
 \cite{sussman2000coupled}. In  order to employ an operator splitting algorithm, the advection of the interface
 (equation.\ref{adveqn1}), using $\nabla \cdot \mathbf{u} = 0$, can be written as:
\begin{equation}
    \frac{\partial C}{\partial t} + \nabla \cdot ( \mathbf{u} C ) = C(\nabla \cdot \mathbf{u}).
    \label{adveqn3}
\end{equation}
The above form of the advection equation is essential for performing volume conserving $x-$direction and $y$-direction sweeps
 separately (see \cite{Pilliod2004}). Given a volume fraction ($C_{i,j}^{n}$) and velocity field
 ($u_{i+1/2,j}^{n},v_{i,j+1/2}^{n}$) at the $nth$ time step, the discretised equation \ref{adveqn3} is given by,
\begin{equation}
    \begin{aligned}
    C_{i, j}^{n+1}= C_{i, j}^{n}+\frac{\Delta t}{r_{i,j} \Delta x}\left(\delta V_{i-1 / 2, j}-\delta V_{i+1 / 2, j}\right) 
        +\frac{\Delta t}{\Delta y}\left(\delta V_{i, j-1 / 2}-\delta V_{i, j+1 / 2}\right) + \\
        C_{i, j}^{n} 
        \left(
        \frac{\Delta t}{r_{i,j} \Delta x} \left( r_{i+1/2,j}u_{i+1/2,j}^{n}-r_{i-1/2,j}u_{i-1/2,j}^{n} \right)  
        + \frac{\Delta t}{\Delta y} \left( v_{i,j+1/2}^{n}-v_{i,j-1/2}^{n} \right)  
        \right)
    \end{aligned}    
    \label{adveqn4}
\end{equation}
where $\delta V_{i+1 / 2, j} = (ruC)_{i+1/2,j}^{n}$ is the amount of volume fraction fluxed through the right cell face. Similarly,
 fluxes $\delta V_{i-1 / 2, j}, \delta V_{i, j+1/2}$ and $\delta V_{i, j-1/2}$ can be computed for other cell faces.

Using operator splitting, we can split the above equation as following:
\begin{equation}
    C_{i, j}^{*}= C_{i, j}^{n}+\frac{\Delta t}{r_{i,j} \Delta x}\left(\delta V_{i-1 / 2, j}-\delta V_{i+1 / 2, j}\right) +
        C_{i, j}^{*} 
        \left(
        \frac{\Delta t}{r_{i,j} \Delta x} \left( r_{i+1/2,j}u_{i+1/2,j}^{n}-r_{i-1/2,j}u_{i-1/2,j}^{n} \right)    
        \right)
    \label{adveqn5}
\end{equation}

\begin{equation}
    C_{i, j}^{n+1}= C_{i, j}^{*} 
        + \frac{\Delta t}{\Delta y}\left(\delta V_{i, j-1 / 2}-\delta V_{i, j+1 / 2}\right) + 
        C_{i, j}^{*} 
        \left(
        \frac{\Delta t}{\Delta y} \left( v_{i,j+1/2}^{n}-v_{i,j-1/2}^{n} \right)  
        \right)
    \label{adveqn6}
\end{equation}
where $C_{i,j}^{*}$ is the intermediate value of the volume fraction. An implicit scheme is used in the first direction and
an explicit scheme in the second direction to maintain the conservation of volume fraction\cite{puckett1997}. The order of
 sweep of direction is alternated every timestep \cite{strang1968}("Strang spliting") to achieve second order accuracy in time.

\begin{figure}[!htbp]
    \centerline{\includegraphics[width=2.5in]{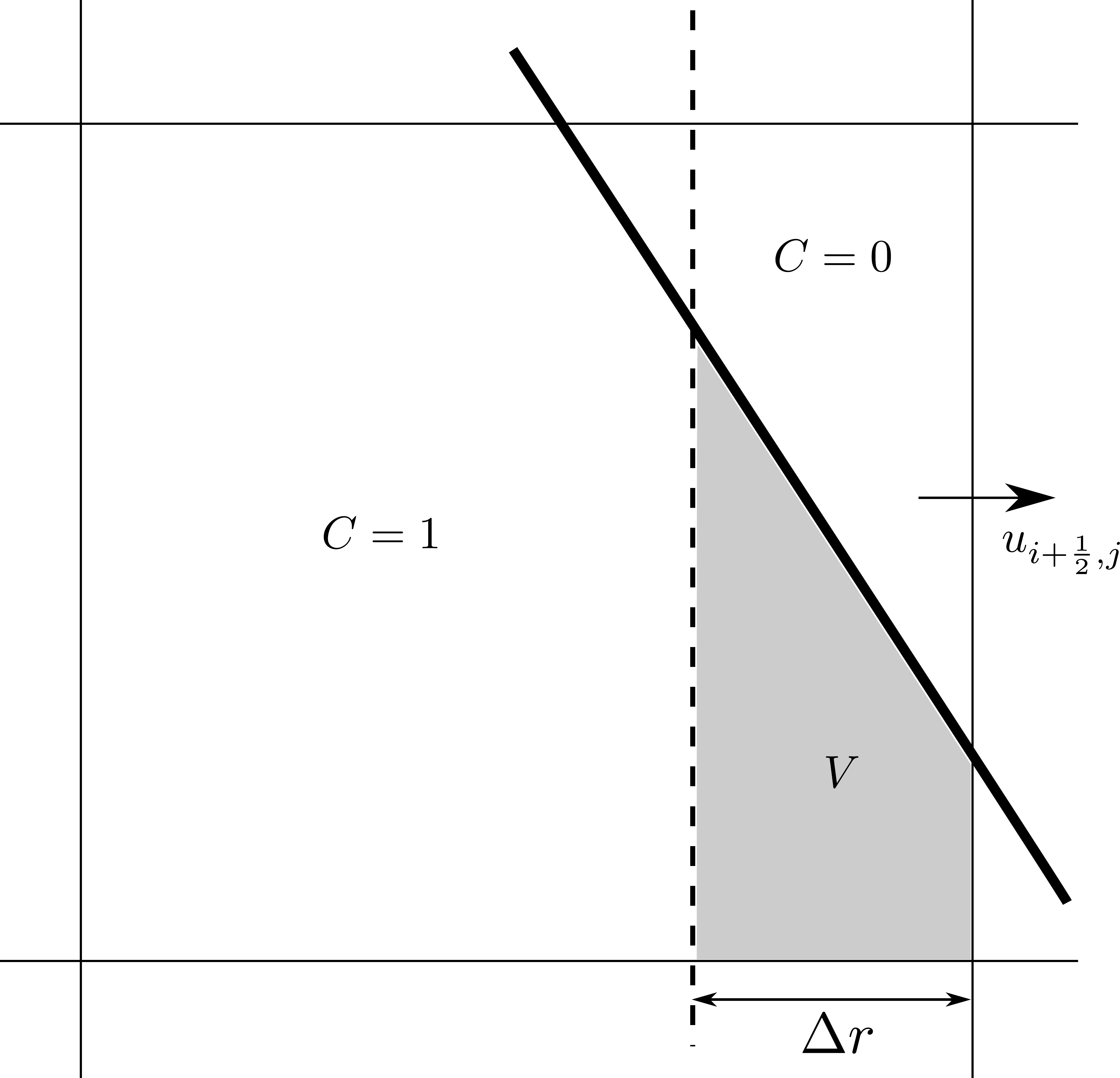}}
    \caption{The fluxed volume through the right face of the cell when $u_{i+1/2,j}$ is positive.}
    \label{flux}
\end{figure}

The volume flux through cell faces, $\delta V_{cell-face}$, is computed geometrically. Consider the schematic in figure
 \ref{flux}, where the shaded region shows the volume of fluid-$1$ in the cell to be fluxed through the right face
 ($\delta V_{i+1/2,j}$). Considering the face velocity ($u_{i+1/2,j}$) to be positive, the flux can be computed as,
\begin{equation}
    \delta V_{i+\frac{1}{2},j} = \frac{(r u)_{i+1/2,j} V}{2 \pi r \Delta r \Delta y}
\end{equation}    
where $V$ is the volume of fluid $1$ fluxed through the right face (shown as the shaded region in figure \ref{flux}), $\Delta r$
is the distance in the radial direction which contains the volume advected in this timestep  and $r$ is
the distance to the center of this volume from the axis of symmetry. We can calculate $\Delta r$ by considering the conservation 
of volume fluxed through the right face and solving the resulting quadratic equation, 
which yields, $\Delta r = r_{i+\frac{1}{2},j}- \sqrt{r_{i+\frac{1}{2},j}^2 - 2 r_{i+\frac{1}{2},j} u_e \Delta t}$. Using the
 section of the piece-wise reconstructed interface lying in the volume to be fluxed through the cell-face over 
$\Delta t$ time-step and employing the Gauss area formula, given by equation \ref{formula_shoelace}, we can calculate the volume
 cut by this region.

\begin{figure}[!htbp]
    \centerline{\includegraphics[width=2.5in]{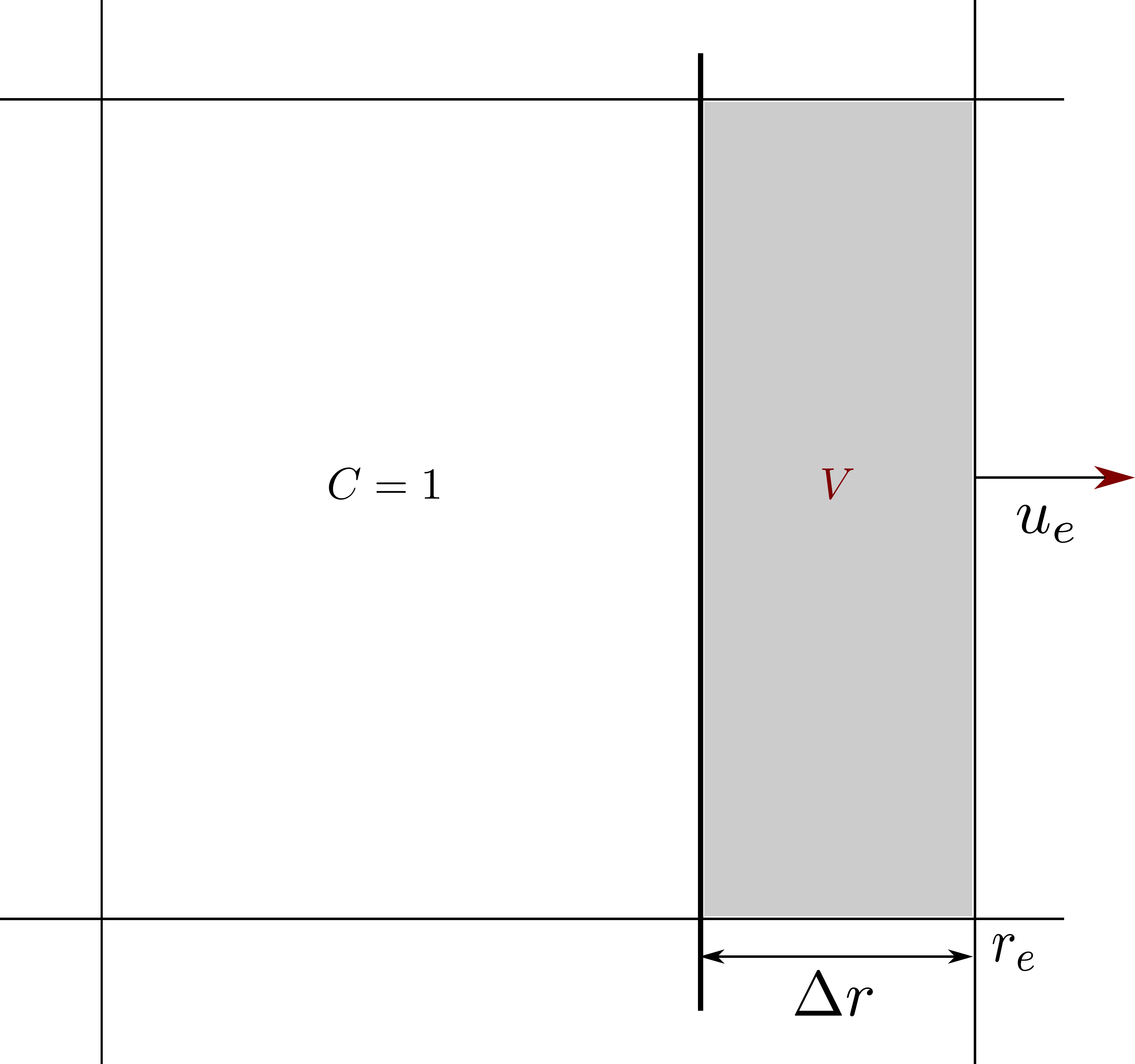}}
    \caption{The shaded region is the volume advected through the east cell face in a single timestep.}
    \label{error}
\end{figure}

The above small correction in computing $\Delta r$ along with the accurate Gauss formula for axisymmetric simulations allows us
 to improve upon the existing volume fraction advection schemes. Existing schemes modify the velocity in $2D$ algorithms by
 using $r\mathbf{u}$ for velocity field and use $2D$ geometric advection scheme which results in a third order error
 ($O(\Delta t^2 h)$, where $h$ is the grid size and $\Delta t$ is the timestep) in mass conservation. We illustrate this by
 considering $x-$direction advection of a small volume of fluid through the right face of the cell with a velocity $u_e$, shown
 as the shaded region in the figure \ref{error} .

The existing schemes compute the volume as $V_c = 2 \pi (r_e - \frac{u_e \Delta t}{2}) (u_e \Delta t) \Delta y$, where
 $r_e$ is the distance of the east cell face from the axis of symmetry, $\Delta t$ is the timestep, and $\Delta y$ is the height
 of the cell. Whereas, the proposed scheme yields the exact volume, $V = 2 \pi (r_e - \frac{\Delta r}{2}) \Delta r \Delta y$ with 
$\Delta r = r_e - \sqrt{r_e^2 - 2 r_e u_e \Delta t}$. Thus, the error in volume calculation is given by, $E = \pi (u_e \Delta t)^2 \Delta y$.

We validate the proposed modifications with the following test cases and compare with the results obtained using the open source
 multiphase flow solver Gerris\cite{popinet2003}.

\subsection{Advection of a torus}

In this test case, a torus of radius $0.25$ is initialised at $(0.35,0.5)$ in a computational domain of size $(2.0,1.0)$. The torus is advected under the steady state velocity of $ u = 0.1/r$ for $r > 0.05$ and $ v = 0 $, where $r$ is the distance from the
 axis of symmetry. The fluid is advected $1000$ timesteps forward in time and then the velocity is reversed to compute $1000$
 timesteps backwards in time. The grid size is $1/128$, the grid Courant number(CFL) is chosen to be $1$ which corresponds to a 
time step of $\Delta t=0.0078125$. As seen from the figure \ref{Test1}, after $1000$ timesteps the torus is highly compressed
 during the advection as less area (due to axisymmetry) occupies the same volume as we move away from the axis of symmetry. We note that the final
 interface shape matches very well with the initial position of the torus, thus validating our algorithm. 

\begin{figure}[!htbp]
    \centerline{\includegraphics[width=5.in]{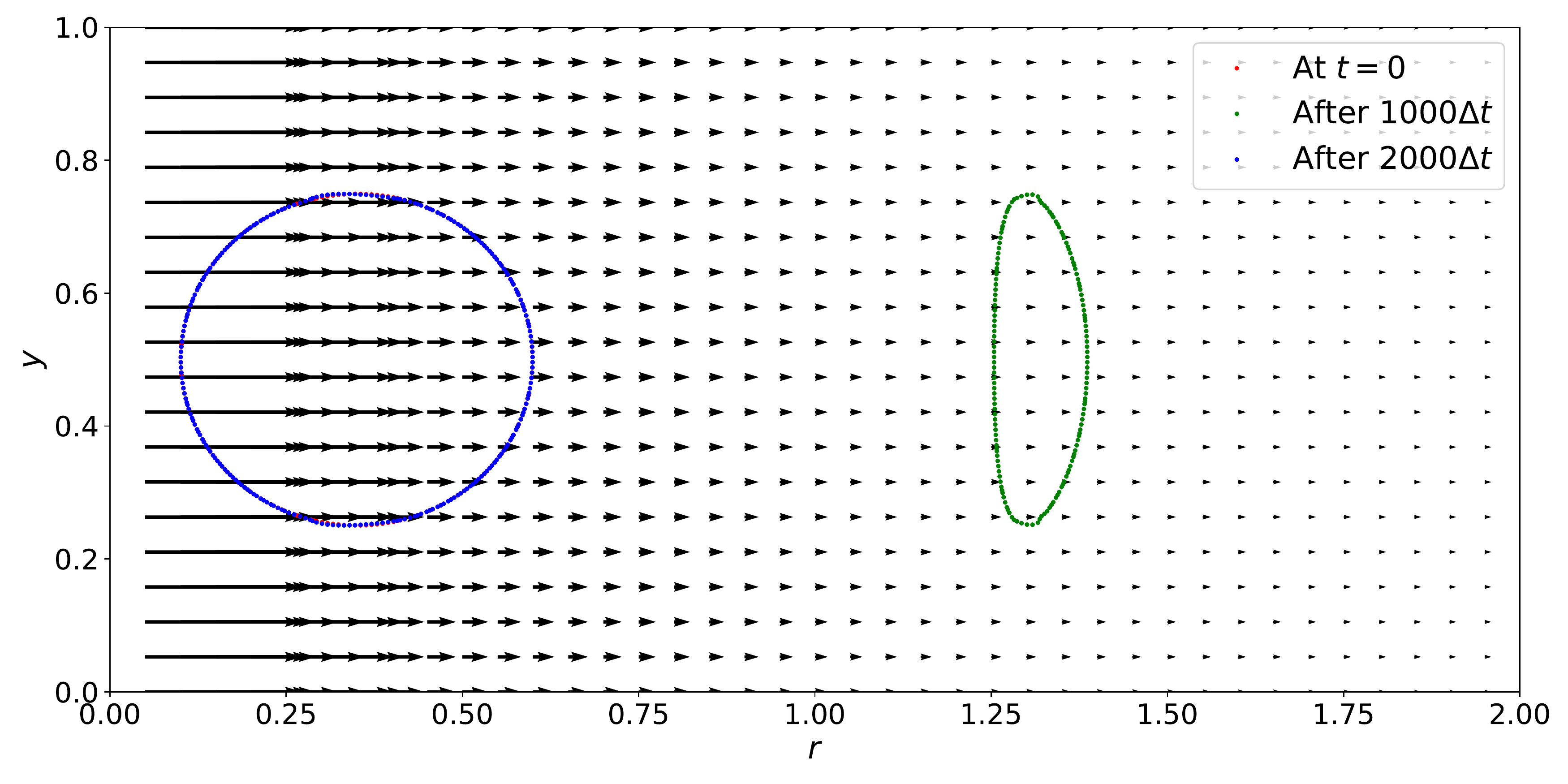}}
    \caption{Interface shape after $1000$ time steps forward and backward after advecting the torus of radius $.25$
    with a grid Courant(CFL) number of $1$.}
    \label{Test1}
\end{figure}

The relative error in the volume between the initial and final distribution of fluid-$1$ for various number of forward and
 backward advection time steps is given in table \ref{table2}. The corresponding relative change in the volume obtained for the same test
 case simulated using Gerris flow solver are also given for comparison. We note that the error obtained from the present schemes are highly
 accurate in comparison to those obtained from Gerris.

\begin{table}
    \caption{Results for relative error in the volume for a torus of radius $0.25$ advected in radial direction forward and
    backward in time for different number of timesteps.}
  \centering
  \begin{tabular}{llll}
    \toprule
    \multicolumn{3}{c}{Relative error in volume}                   \\
    \cmidrule(r){1-3}
 Number of timesteps& Current Solver&Gerris Solver\\
    \midrule
    $1$   &       $ \num{5.2e-15}$     & $ \num{1.1e-5}$\\
    $10$   &      $ \num{2.9e-14}$     & $\num{7.2e-5}$\\
    $100$   &     $ \num{4.8e-13}$     & $\num{2.1e-4}$\\
    $1000$   &    $ \num{3.4e-10}$    & $\num{3.8e-4}$\\
    \bottomrule
  \end{tabular}
    \label{table3}
\end{table}


 As suggested by Kothe et al.\cite{rider1995}, simple linear advection test cases do not reveal the efficacy of advection
 algorithms appropriately. Thus, we further test the efficacy of the algorithm by subjecting it to a more severe test case of
 advection of a torus in a Hill's vortex. This is axisymmetric equivalent of the circle in a vortex test case for 2D cases
 \cite{rider1995}. For this velocity field,  the interface undergoes strong topological changes including fragmentation
 and merging due to strong shear effects. Here we use a modified form of Hills's vortex with a superimposed radial flow field.
 A torus of radius $0.1$ is initialised at $(0.2,0.8)$ in a computational domain of size $(1.0,1.0)$ with $L=0.5$. The fluid is
advected under highly strained steady state velocity field given by
\begin{eqnarray}   
  u =&0.1\left(\frac{r}{L} \frac{(y-L)}{L}\right) + \frac{0.05}{r}\\
    v=&0.1\left[1-\left(\frac{y-L}{L}\right)^{2}-2\left(\frac{r}{L}\right)^{2}\right].
\end{eqnarray}

The fluid is advected $4000$ timesteps forward in time and then the velocity is reversed to advect $4000$ timesteps backwards
 in time. The grid size is $1/128$ and the time step is $\num{1.0e-3}$.

\begin{figure}[!htbp]
    \centerline{\includegraphics[width=3.0in]{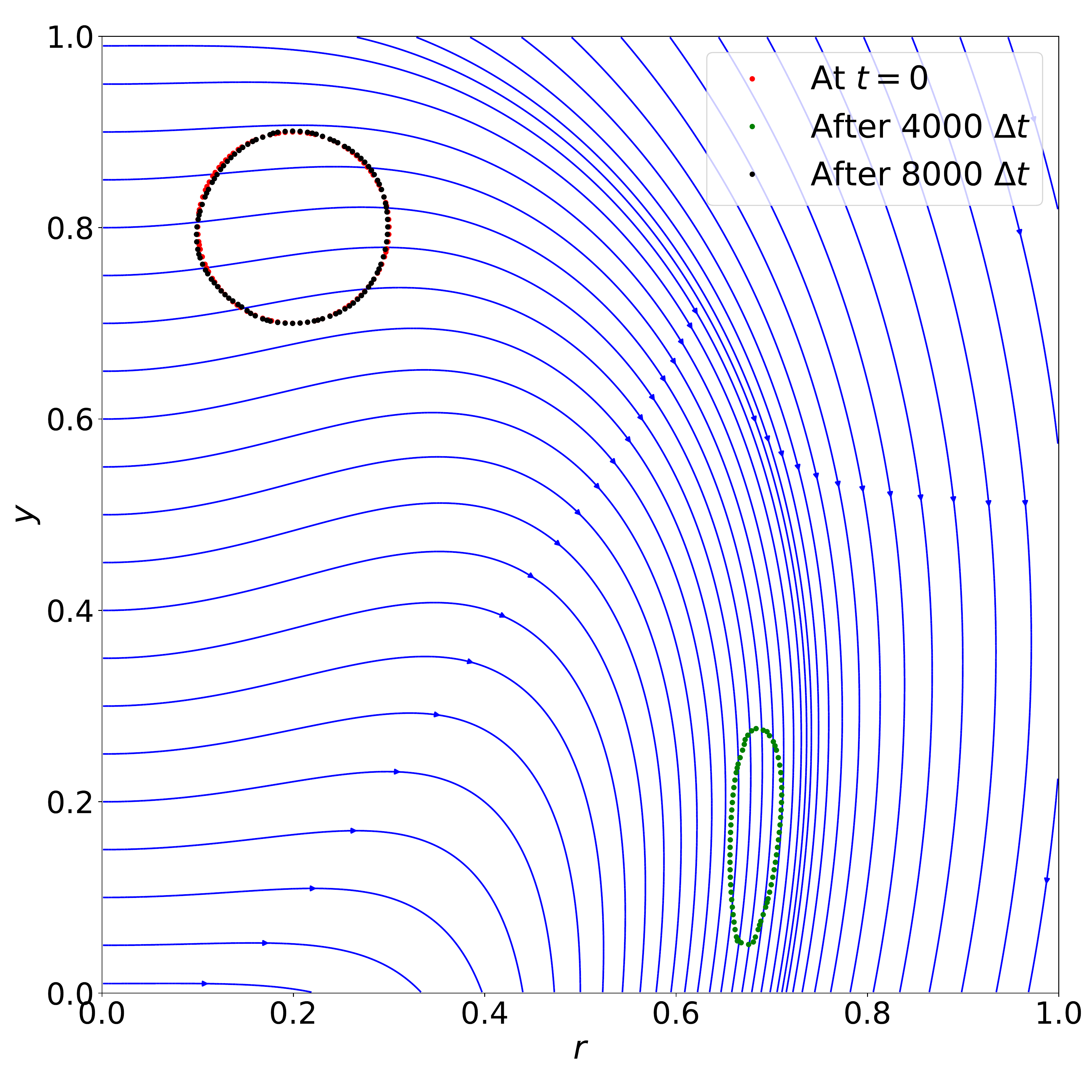}}
    \caption{Interface shape after $4000$ time steps forward and backward after advecting the torus placed in a vortex. 
    The relative error in change in volume is $\num{1.7e-6}$}
    \label{Test3}
\end{figure}

As seen from figure \ref{Test3} after $4000$ timesteps the shape of the interface is highly distorted due to highly strained
 velocity field. The final interface shape matches very well with the initial position of the toroid thus validating our
 algorithm. The relative error in change in volume between the initial and the final distribution of fluid $1$ for various number
 on time steps is given in table \ref{table3}. The corresponding relative change in volume for the same test case in Gerris flow
 solver. We note the error in the proposed scheme, even for larger number of timesteps, is an order smaller compared to the
 results from Gerris flow solver.

\begin{table}
    \caption{Results for relative error in volume for a torus of radius $0.1$ placed in a complex velocity field advected forward
    and backward in time for different number of timesteps.}
  \centering
  \begin{tabular}{llll}
    \toprule
    \multicolumn{3}{c}{Relative error in volume}                   \\
    \cmidrule(r){1-3}
 Number of timesteps& Current Solver&Gerris Solver\\
    \midrule
    $1$   &       $ \num{2.7e-10}$     & $ \num{2.2e-6}$\\
    $10$   &      $ \num{5.2e-10}$     & $\num{4.5e-6}$\\
    $100$   &     $ \num{2.5e-8}$     & $\num{1.8e-5}$\\
    $1000$   &    $ \num{1.3e-6}$    & $\num{4.5e-5}$\\
    $4000$   &    $ \num{1.7e-6}$    & $\num{4.8e-5}$\\
    \bottomrule
  \end{tabular}
    \label{table4}
\end{table}

%

\subsection{Bubble in a Stagnation Point Flow}
In this test case we implement the VOF algorithm presented in this paper for a more complex flow. We solve Navier-Stokes 
equations in one fluid form given by:
\begin{equation}
\rho(C) \left( \frac{\partial \mathbf{u}}{\partial t} +  \mathbf{u} . {\nabla} \mathbf{u} \right) = -{\nabla} p + {\nabla} \cdot \big[\mu(C) \left( {\nabla} \mathbf{u} + ({\nabla}\mathbf{u})^{\small T} \right)\big] + \rho(C) \mathbf{g} + \mathbf{f}^{\gamma}_{v}
\label{one_fluid_form}
\end{equation}
where $\mathbf{u}$ and $p$ are the velocity vector and pressure, respectively, $\rho(C)$  and $\mu(C)$ are the fluid density and
 viscosity which are a function of void fraction field, $C$. We use Chorin's projection method~\cite{chorin1968numerical} to 
solve the above equation \ref{one_fluid_form} where we discretise the advection term using a second order ENO
 scheme~\cite{shu1988efficient} and the diffusion terms using central differencing. Surface tension forces,
 $\mathbf{f}^{\gamma}_{v}$ are acting only at the interface and have been modeled as volumetric body force using the continuum
 surface force model of Brackbill, Kothe, and Zemach~\cite{brackbill1992continuum}. The interface is  captured using CLSVOF
 algorithm given by Sussman and Puckett~\cite{sussman2000coupled}. This algorithm is mass conserving and
 calculates the curvature and surface normal with high accuracy which is used for surface tension force calculation.
 The interface  is advected by solving the advection equations for level-set function, $\phi$, and volume fraction, $C$.
\begin{figure}[!htbp]
    \centerline{\includegraphics[width=4in]{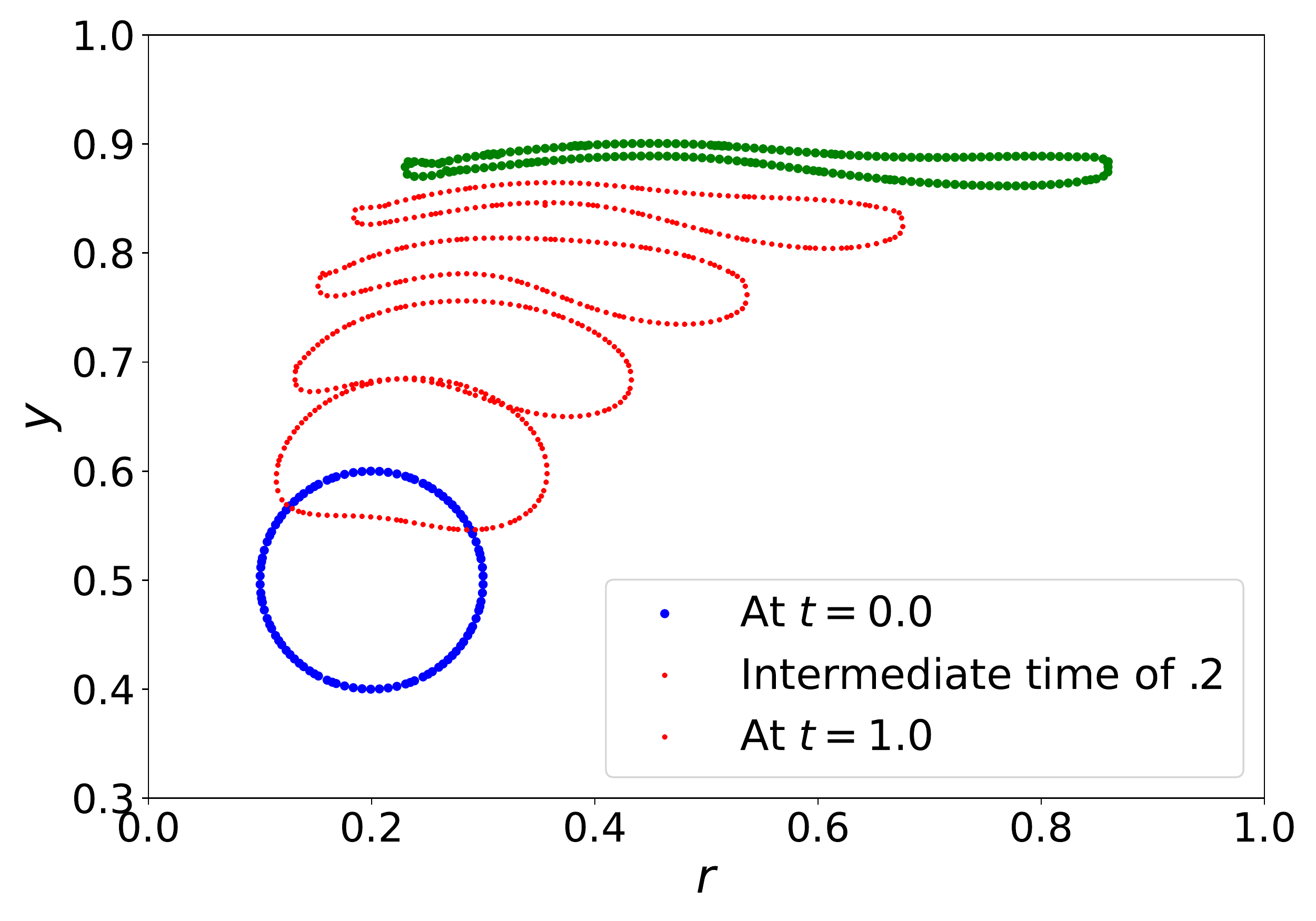}}
    \caption{Interface shape after time $t=1$. The interface is flattened against the top wall and stretched due to the underlying
    velocity. Even when the cross-sectional area changes the volume of the toroidal bubble is maintained very accurately.}
    \label{Test4}
\end{figure}
We initialize a toroidal bubble of radius $0.1$ at $(0.2,0.5)$ in a computational domain of unit size, $1 \times 1$. The
bottom boundary has an inlet velocity of unity in the upward axial direction and the right boundary has outflow boundary 
conditions. The top boundary acts as a rigid wall with no-slip and impermeable surface. The density ratio and viscosity
 ratio is $10$ with the Laplace number of the bubble, $La = \frac{\rho D \sigma}{\mu^2} = 0.048$. The incoming axial velocity
 drags the bubble and flattens it against the top wall, stretching it in the axial direction considerably. Even though the
 fluid interface undergoes drastic change in its shape, the volume is conserved with a high degree of accuracy with
 relative error in the volume of \num{2.1e-6}.

\section{Conclusions}

In the present work, we have presented several improvements for the implementation of volume of fluid method in axisymmetric
 coordinates. We have presented analytical relations for the reconstruction of piecewise linear interface in axisymmetric
 coordinates similar to those given by Scardovelli and Zaleski\cite{Scardovelli2000} for cartesian coordinates. The
 proposed scheme substantially reduces the computational cost in comparison to the iterative schemes usually employed for
 the reconstruction. Further, we showed that even for axisymmetric coordinate system, machine-precision advection of volume
 fraction field can be achieved. We illustrated the improvements by comparing the results with the popular open source
 multiphase flow solver Gerris. Finally, we would like to note that similar modifications in the advection scheme for  volume of fluid method in
 other curvillinear coordinate systems(such as elliptic coordinates) can be derived using the approach presented in this work.

\bibliographystyle{unsrt}  
\bibliography{references}  

\end{document}